\documentclass{ws-procs9x6}
\usepackage{epsfig}

\def\beq{\begin{equation}}
\def\eeq{\end{equation}}
\def\bea{\begin{eqnarray}}
\def\eea{\end{eqnarray}}
\def\cal{}

\begin{document}

\title{Lifetimes of heavy 
hadrons\footnote{\uppercase{T}his work is supported by the
\uppercase{U.S.\ N}ational \uppercase{S}cience \uppercase{F}oundation
under \uppercase{G}rant \uppercase{PHY}--0244853, and by the 
\uppercase{U.S.\ D}epartment of \uppercase{E}nergy under
\uppercase{C}ontract \uppercase{DE-FG02-96ER41005}.
}}

\author{ALEXEY A. PETROV}

\address{Department of Physics and Astronomy\\
Wayne State University\\ 
Detroit, MI 48201, USA\\ 
E-mail: apetrov@physics.wayne.edu}

\maketitle

\abstracts{
We review current status of theoretical predictions of lifetimes of 
heavy hadrons in heavy-quark expansion. We present a calculation of 
subleading $1/m_b$ corrections to spectator effects
in the ratios of beauty hadron lifetimes. We find that
these effects are sizable and should be taken into account in systematic analyses
of heavy hadron lifetimes. In particular, the inclusion of $1/m_b$ corrections
brings into agreement the theoretical predictions and experimental observations of
the ratio of lifetimes of $\Lambda_b$-baryon and $B_d$ meson. We
obtain $\tau(B_u)/\tau(B_d) = 1.09 \pm 0.03, \,
\tau(B_s)/\tau(B_d)= 1.00 \pm 0.01, \, \tau(\Lambda_b)/\tau(B_d)= 0.87
\pm 0.05$.
}

\section{Introduction}

The hierarchy of lifetimes of heavy hadrons can be understood in
the heavy-quark expansion (HQE), which makes use of the disparity of scales present in the
decays of hadrons containing b-quarks. HQE predicts the ratios of lifetimes of beauty
mesons\cite{Bigi:1994wa,Neubert:1996we,Rosner:1996fy}, which agree with the 
experimental observations well within experimental and {\it theoretical} 
uncertainties. Most recent experimental analyses give\cite{PDG,ave}
\bea
\tau(B_u)/\tau(B_d)|_{ex}&=&1.085 \pm0.017, \nonumber \\
\tau(B_s)/\tau(B_d)|_{ex}&=&0.951 \pm 0.038, \\
\tau(\Lambda_b)/\tau(B_d)|_{ex}&=& 0.797 \pm 0.053. \nonumber
\eea
The most recent theoretical predictions show evidence of excellent agreement of theoretical 
and experimental results\cite{Ciuchini:2001vx,Gabbiani:2003pq}. 
This agreement also provides us with some confidence that quark-hadron
duality, which states that smeared partonic amplitudes can be replaced by the hadronic
ones, is expected to hold in inclusive decays of heavy flavors. It should be pointed out 
that the low experimental value of the ratio $\tau(\Lambda_b)/\tau(B_d)$ has long been a 
puzzle for the theory. Only recent next-to-leading order (NLO) calculations of
perturbative QCD\cite{Ciuchini:2001vx} and $1/m_b$ corrections\cite{Gabbiani:2003pq}
to spectator effects significantly reduced this discrepancy. Of course, the problem of 
$\tau(\Lambda_b)/\tau(B_d)$ ratio could reappear again if future measurements at Fermilab 
and CERN would find the mean value to stay the same with error bars shrinking. 
Upcoming Fermilab measurements of $\Lambda_b$ lifetime could shed more light on the 
experimental side of this issue.

This talk reports on the calculation of subleading contributions to spectator effects 
in the $1/m_b$ expansion\cite{Gabbiani:2003pq} to study their impact on the ratios of 
lifetimes of heavy mesons. We also discuss the convergence of the $1/m_b$ expansion in 
the analysis of spectator effects.

\section{Formalism}\label{formalism}

The inclusive decay rate of a heavy hadron $H_b$ is most conveniently computed by employing
the optical theorem to relate the decay width to the imaginary part of the forward matrix
element of the transition operator:
\beq \label{rate}
\Gamma(H_b)=\frac{1}{2 M_{H_b}} \langle H_b |{\cal T} | H_b \rangle,~~
{\cal T} = {\mbox{Im}}~ i \int d^4 x T \left\{
H_{\mbox{\scriptsize eff}}(x) H_{\mbox{\scriptsize eff}}(0) \right \}.
\eeq
Here $H_{\mbox{\scriptsize eff}}$ represents an effective $\Delta B=1$ Hamiltonian,
\beq
H_{\mbox{\scriptsize eff}} = \frac{4 G_F}{\sqrt{2}} V_{cb} \sum_{d'=d,s, u'=u,c}
V^*_{u'd'} \left[ C_1(\mu) Q_1^{u'd'} (\mu) +
C_2(\mu) Q_2^{u'd'} (\mu) \right] + h.c.,
\eeq
where the four-quark operators $Q_1$ and $Q_2$ are given by
\beq
Q_1^{u'd'}=\bar d_{L}' \gamma_\mu u_{L}' ~\bar c_{L} \gamma^\mu b_{L},\qquad
Q_2^{u'd'}=\bar c_{L} \gamma_\mu u_{L}' ~\bar d_{L}' \gamma^\mu b_{L}.
\eeq
In the heavy-quark limit, the energy release is large, so the correlator in
Eq.~(\ref{rate}) is dominated by short-distance physics.
An Operator Product Expansion (OPE) can be constructed for Eq.~(\ref{rate}),
which results in a prediction of decay widths of Eq.~(\ref{rate}) as
a series of local operators of increasing dimension suppressed by powers
of $1/m_b$:
\beq\label{expan}
\Gamma(H_b)= \frac{1}{2 M_{H_b}} \sum_k \langle H_b |{\cal T}_k | H_b \rangle
=\sum_{k} \frac{C_k(\mu)}{m_b^{k}}
\langle H_b |{\cal O}_k^{\Delta B=0}(\mu) | H_b \rangle.
\eeq
In other words, the calculation of $\Gamma(H_b)$ is equivalent to computing
the matching coefficients of the effective $\Delta B=0$ Lagrangian
with subsequent computation of its matrix elements. Indeed, at the end, the
scale dependence of the Wilson coefficients in Eq.~(\ref{expan}) should match
the scale dependence of the computed matrix elements.

It is customary to make predictions for the ratios of lifetimes (widths),
as many theoretical uncertainties cancel out in the ratio. In addition, since
the differences of lifetimes should come from the differences in the ``brown mucks''
of heavy hadrons, at the leading order in HQE all beauty hadrons with light
spectators have the same lifetime. Thus, all ratios converge to unity in the 
heavy quark limit.

The difference between meson and baryon lifetimes first occurs at order
$1/m^2$ and is essentially due to the different structure of mesons and
baryons. In other words, no lifetime difference is induced among the members
of the meson multiplet at this order. The ratio of heavy meson and baryon lifetimes
receives a shift which amounts to at most $1-2\%$, not sufficient
to explain the observed pattern of lifetimes\cite{Neubert:1996we}.

The main effect appears at the $1/m^3$ level and comes from the set of
dimension-six four-quark operators, whose contribution is
enhanced due to the phase-space factor $16 \pi^2$. They are thus capable of inducing
corrections of order $16 \pi^2 (\Lambda_{QCD}/m_b)^3$ = ${\cal O}(5-10\%)$.
These operators, which are commonly called Weak Annihilation (WA) and Pauli Interference
(PI), introduce a difference in lifetimes for both
heavy mesons and baryons. Their effects have been computed\cite{Neubert:1996we,Guberina:1979xw}
at leading order in perturbative QCD, and, more recently, including NLO perturbative
QCD corrections\cite{Ciuchini:2001vx} and $1/m_b$ corrections\cite{Gabbiani:2003pq}.
The contribution of these operators to the lifetime ratios are
governed by the matrix elements of $\Delta B=0$ four-fermion operators
\bea
{\cal T}_{\rm spec} &=&
{\cal T}_{\rm spec}^{u} +
{\cal T}_{\rm spec}^{d'} +
{\cal T}_{\rm spec}^{s'},
\eea
where the ${\cal T}_{i}$ are
\bea\label{SpecLO}
{\cal T}_{\rm spec}^{u} &=&
\frac{G_F^2m_b^2|V_{bc}|^2(1-z)^2}{2\pi}
\Big\{
\left(c_1^2+c_2^2\right)
O_1^{u}
+2c_1c_2\widetilde{O}_1^{u}
+ \delta_{1/m}^{u}
\Big\}, \nonumber \\
{\cal T}_{\rm spec}^{d'} &=&
-\frac{G_F^2m_b^2|V_{bc}|^2(1-z)^2}{4\pi}
\left\{
c_1^2\left[
(1+z)O_1^{d'}+\frac{2}{3}(1+2z)O_2^{d'}
\right] \right. \nonumber \\ && \left.
+\left(
N_cc_2^2+2c_1c_2
\right)
\left[
(1+z)\widetilde{O}_1^{d'}+\frac{2}{3}(1+2z)\widetilde{O}_2^{d'}
\right]
+\delta_{1/m}^{d'}
\right\}, \\
{\cal T}_{\rm spec}^{s'} &=&
-\frac{G_F^2m_b^2|V_{bc}|^2\sqrt{1-4z}}{4\pi}
\left\{
c_1^2\left[
O_1^{s'}+\frac{2}{3}(1+2z)O_2^{s'}
\right] \right. \nonumber \\ && \left.
+\left(N_cc_2^2+2c_1c_2
\right)
\left[
\widetilde{O}_1^{s'}+\frac{2}{3}(1+2z)\widetilde{O}_2^{s'}
\right]
+\delta_{1/m}^{s'}
\right\}. \nonumber
\eea
Here the terms $\delta^i_{1/m}$ refer to $1/m_b$
corrections to spectator effects, which we discuss below. Note that we include
the full $z = {m_c^2}/{m_b^2}$ dependence, which is fully consistent only after the
inclusion of higher $1/m_b$ corrections. The operators $O_i$ and $\widetilde{O}_i$ in
Eq.~(\ref{SpecLO}) are defined as
\bea
O_1^q = \bar b_i\gamma^{\mu}(1-\gamma_5)b_i\bar q_j\gamma_{\mu}(1-\gamma_5)q_j, &&
O_2^q = \bar b_i\gamma^{\mu}\gamma_5b_i\bar
q_j\gamma_{\mu}(1-\gamma_5)q_j,\nonumber \\
\widetilde{O}_1^q = \bar b_i\gamma^{\mu}(1-\gamma_5)b_j
\bar q_i\gamma_{\mu}(1-\gamma_5)q_j, &&
\widetilde{O}_2^q = \bar b_i\gamma^{\mu}\gamma_5 b_j
\bar q_i\gamma_{\mu}(1-\gamma_5)q_j.
\eea
In order to assess the impact of these and other operators, parameterizations of
their matrix elements must be introduced. The meson matrix elements are
\bea \label{MEM}
\langle B_q |O_1^q| B_q \rangle &=& f^2_{B_q} m^2_{B_q} \left(2 \epsilon_1 + \frac{B_1}{N_c}
\right), ~~
\langle B_q |\widetilde{O}_1^q| B_q \rangle = f^2_{B_q} m^2_{B_q} B_1,
\nonumber \\
\langle B_q |O_2^q| B_q \rangle &=& -f^2_{B_q} m^2_{B_q} \left[
\frac{m_{B_q}^2}{\left(m_b+m_q\right)^2} \left(2 \epsilon_2 + \frac{B_2}{N_c}\right)
+ \frac{1}{2}\left(2 \epsilon_1 + \frac{B_1}{N_c}
\right)
\right],~~~ \\
\langle B_q |\widetilde{O}_2^q| B_q \rangle &=& -f^2_{B_q} m^2_{B_q} \left[
\frac{m_{B_q}^2}{\left(m_b+m_q\right)^2} B_2 + \frac{1}{2} B_1
\right]. \nonumber
\eea
Here the parameters $B_i$ and $\epsilon_i$ are usually referred to as
``singlet'' and ``octet'' bag parameters\cite{Neubert:1996we}.
Expressed in terms of these parameters, the lifetime ratios of heavy mesons
can be written as
\bea \label{MesonMeson}
\tau(B_u)/\tau(B_d)= 1+16 \pi^2 \frac{f_B^2 m_B}{m_b^3 c_3(m_b)}
\left[G_1^{ss} (m_b) B_1 (m_b)+G_1^{oo} (m_b) \epsilon_1 (m_b)
\right.
\nonumber \\
\left. + ~G_2^{ss} (m_b) B_2 (m_b)+ G_1^{oo} (m_b) \epsilon_2 (m_b) \right] 
+ \delta_{1/m},~~~
\eea
where the coefficients $G$ were computed at NLO\cite{Ciuchini:2001vx}.
$\delta_{1/m}$ represents spectator corrections of order $1/m_b$ and higher,
which we discuss below.

Calculations of the matrix elements of four-fermion operators in baryon decays are
not straightforward. Similar to the meson matrix elements above, we 
relate them to the value of $\Lambda_b$-baryon wave function at the origin,
\bea
\langle \Lambda_b |O_1^q| \Lambda_b \rangle &=&
-\widetilde{B} \langle \Lambda_b |\widetilde{O}_1^q| \Lambda_b \rangle =
\frac{\widetilde{B}}{6}  f^2_{B_q} m_{B_q} m_{\Lambda_b} r, \quad
\nonumber \\
\langle \Lambda_b |O_2^q| \Lambda_b \rangle &=&
-\widetilde{B} \langle \Lambda_b |\widetilde{O}_2^q| \Lambda_b \rangle=
\frac{\widetilde{B}}{6} f^2_{B_q} m_{B_q} m_{\Lambda_b} \delta.
\eea
Here the parameter $\widetilde{B}$ accounts for the deviation of the 
$\Lambda_b$ wave function from being totally color-asymmetric\cite{Neubert:1996we}
($\widetilde{B}=1$ in the valence approximation), and the parameter
$r=\left|\psi^{\Lambda_b}_{bq}(0)\right|^2/\left|\psi^{B_q}_{b\bar q}(0)\right|^2$
is the ratio of the wave functions at the origin of the $\Lambda_b$ and $B_q$ mesons.
Note that $\delta={\cal O}(1/m_b)$, which follows from the heavy-quark spin
symmetry. It needs to be included as we consider higher-order corrections
in $1/m_b$. 
While these parameters have not been computed model-independently,
various quark-model arguments suggest that the meson and baryon matrix elements
are quite different. Thus a meson-baryon lifetime difference can be
produced. In general, one can parametrize the meson-baryon lifetime
ratio as
\bea \label{MesonBaryon}
\tau(\Lambda_b)/\tau(B_d) \simeq  0.98 - (d_1+d_2 \widetilde{B})r
-(d_3 \epsilon_1+d_4 \epsilon_2) -(d_5 B_1 + d_6 B_2) \nonumber \\
 \simeq  0.98 - m_b^2 (d_1'+d_2' \widetilde{B})r
-m_b^2 \left[(d_3' \epsilon_1+d_4' \epsilon_2) +(d_5' B_1 + d_6' B_2)\right],~
\eea
where in the last line we scaled out the coefficient $m_b^2$ emphasizing
the fact that these corrections are suppressed by
$1/m_b^3$ compared to the leading $m_b^5$ effect. The scale-dependent
parameters are $d_i(m_b)$ =  \{0.023, 0.028, 0.16, --0.16, 0.08, --0.08\} at 
NLO\cite{Neubert:1996we}.

It is interesting to note that in the absence of $1/m_b$ corrections to spectator
effects, it would be equally correct to substitute the $b$-quark mass in
Eq.~(\ref{MesonBaryon}) with the corresponding meson and baryon masses, so
\begin{eqnarray}
\tau(\Lambda_b)/\tau(B_d) \simeq
0.98 &-& m_{\Lambda_b}^2 (d_1'+d_2' \widetilde{B})r \nonumber \\
&-&m_{B_d}^2 \left[(d_3' \epsilon_1+d_4' \epsilon_2) +(d_5' B_1 + d_6' B_2)\right],
\end{eqnarray}
which reflects the fact that WS and PI effects occur for the heavy and light quarks
initially bound in the $B_d$ meson and $\Lambda_b$ baryon, respectively. While correct
up to the order $1/m_b^3$, these simple substitutions reduce the ratio of lifetimes
by approximately 3-4\%! We take this as an indication of the importance of bound-state
effects on the spectator corrections, represented by subleading $1/m_b$ corrections
to spectator operators.

\section{Subleading corrections to spectator effects}\label{AllResults}

We computed the higher order corrections, including charm quark-mass effects,
to Eq.~(\ref{SpecLO}) in the heavy-quark expansion, denoted below as
$\delta_{1/m}^{q}$.

\begin{figure}[tb]
\centerline{\epsfxsize=7.5cm\epsffile{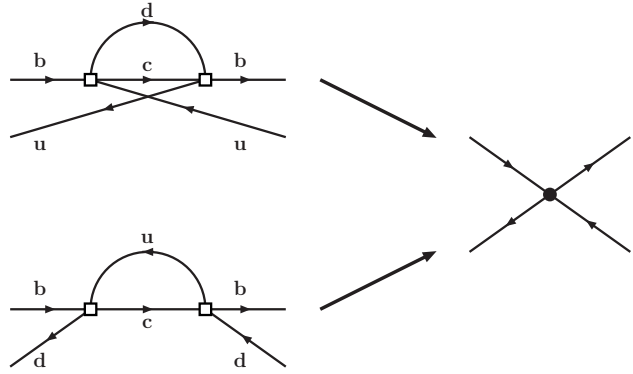}}
\centerline{\parbox{11cm}{\caption{\label{fig:WA}
Kinetic corrections to spectator effects. The operators of Eqs.~(\ref{OurCorrection})
are obtained by expanding the diagrams in powers of
spectator's momentum.}}}
\end{figure}

The $1/m_b$ corrections to the spectator effects were computed\cite{Gabbiani:2003pq}
by expanding the forward scattering amplitude of Eq.~(\ref{rate}) in the light-quark momentum
and matching the result onto the four-quark operators containing derivative insertions
(see Fig.~\ref{fig:WA}). The resulting $\delta_{1/m}^{q}$ contributions can be written in
the following form:
\bea\label{OurCorrection}
&& \delta_{1/m}^{u} =
-2\left(c_1^2+c_2^2\right)\frac{1+z}{1-z}R_1^{u}
-4c_1c_2\frac{1+z}{1-z}\widetilde{R}_1^{u}, \nonumber \\
&& \delta_{1/m}^{d'} =
c_1^2\left[
\frac{8z^2}{1-z}R_0^{d'}
+\frac{2}{3}\frac{1+z+10z^2}{1-z}R_1^{d'}
+\frac{2}{3}(1+2z)\left(
R_2^{d'}-R_3^{d'}
\right)
\right], \nonumber \\ 
&&~+\left(N_cc_2^2+2c_1c_2\right)
\left[
\frac{8z^2}{1-z}\widetilde{R}_0^{d'}
+\frac{2}{3}\frac{1+z+10z^2}{1-z}\widetilde{R}_1^{d'}
+\frac{2}{3}(1+2z)\left(
\widetilde{R}_2^{d'}-\widetilde{R}_3^{d'}
\right)
\right] \nonumber \\
&& \delta_{1/m}^{s'} =
c_1^2\left[
\frac{16z^2}{1-4z}R_0^{s'}
+\frac{2}{3}\frac{1-2z+16z^2}{1-4z}R_1^{s'}
+\frac{2}{3}(1+2z)
\left(
R_2^{s'}-R_3^{s'}
\right)
\right] \\ 
&&~+\left(N_cc_2^2+2c_1c_2\right)
\left[
\frac{16z^2}{1-4z}\widetilde{R}_0^{s'}
+\frac{2}{3}\frac{1-2z+16z^2}{1-4z}\widetilde{R}_1^{s'}
+\frac{2}{3}(1+2z)
\left(
\widetilde{R}_2^{s'}-\widetilde{R}_3^{s'}
\right)
\right].\nonumber
\eea
where the following operators contribute
\bea\label{Operators}
R_0^q &=& \frac{1}{m_b^2}\bar b_i\gamma^{\mu}\gamma_5\vec{D}^{\alpha}b_i
\bar q_j\gamma_{\mu}(1-\gamma_5)\vec{D}_{\alpha}q_j, \nonumber \\
R_1^q &=& \frac{1}{m_b^2}\bar b_i\gamma^{\mu}(1-\gamma_5)\vec{D}^{\alpha}b_i
\bar q_j\gamma_{\mu}(1-\gamma_5)\vec{D}_{\alpha}q_j,  \nonumber \\
R_2^q &=& \frac{1}{m_b^2}\bar b_i\gamma^{\mu}(1-\gamma_5)\vec{D}^{\nu}b_i
\bar q_j\gamma_{\nu}(1-\gamma_5)\vec{D}_{\mu}q_j, \\
R_3^q &=& \frac{m_q}{m_b}\bar b_i(1-\gamma_5)b_i\bar q_j(1-\gamma_5)q_j.
\nonumber
\eea
Here $\widetilde{R}_i^q$ denote the color-rearranged
operators that follow from the expressions for $R_i^q$ by
interchanging color indexes of $b_i$ and $q_j$ Dirac spinors.
Note that the above result contains {\it full} QCD $b$-fields, thus there
is no immediate power counting available for these operators. The power
counting becomes manifest at the level of the matrix elements\cite{Beneke:1996gn}.

In order to include above corrections into the prediction of lifetime 
ratios a calculation of meson and baryon matrix elements of the operators
in Eq.~(\ref{Operators}) must be performed. We use factorization to 
guide our parameterizations of $\Lambda_b$ and meson matrix elements, but 
keep matrix elements which vanish in factorization. This is important, as 
the Wilson coefficients of these operators are larger than the ones 
multiplying the operators whose matrix elements survive in the $N_c \to \infty$ limit.

Our parameterizations for meson and baryon matrix elements can be found
in\cite{Gabbiani:2003pq}, where it was shown that
a set of $1/m_b$-corrections to spectators effects can be
parametrized by eight new parameters $\beta_i$ and
$\widetilde{\beta}_i$ ($i=1,...,4$) for heavy mesons and eight new
parameters $\beta_i^\Lambda$ and $\widetilde{\beta}_i^\Lambda$ for
heavy baryons. Although model-independent values of these parameters
will not be known until dedicated lattice simulations are performed,
we presented an estimate of these parameters based on quark model
arguments. In our numerical results we assume the value of the $b$-quark 
pole mass to be $m_b=4.8\pm0.1$~GeV and $f_B=200\pm25$~MeV, as well as
lattice-inspired values of $B_i$ and $\epsilon_i$ parameters\cite{Chay:1999pa}.

Numerically, the set of $1/m_b$ corrections does not markedly affect the ratios of
meson lifetimes, changing the $\tau(B_u)/\tau(B_d)$ and the $\tau(B_s)/\tau(B_d)$ 
ratios by less than half a percent. The effect is more pronounced in
the ratio of $\Lambda_b$ and $B_d$ lifetimes, where it constitutes a $40-45\%$ of the
leading spectator (WA plus PI) contribution, or an overall correction of about $-3\%$
to the $\tau(\Lambda_b)/\tau(B_d)$ ratio. While such a sizable effect is
surprising, the main source of such a large correction can be readily identified.
While the individual $1/m_b$ corrections to WS and PI are of order $20\%$, as expected
from the naive power counting, they contribute to the $\Lambda_b$ lifetime with the
same (negative) sign, instead of destructively interfering as do WS and
PI\cite{Rosner:1996fy,Gabbiani:2003pq}. This conspiracy of two $\sim 20\%$
effects produces such a sizable shift in the ratio of the $\Lambda_b$ and 
$B$-meson lifetimes.

Since all three heavy mesons belong to the same $SU(3)$ triplet, their lifetimes
are the same at order $1/m_b^2$. The computation of the ratios of heavy meson
lifetimes is equivalent to the computation of $U$-spin or isospin-violating
corrections. Both $1/m_b^3$-suppressed spectator effects and our
corrections computed in the previous sections arise from the spectator interactions
and thus provide a source of $U$-spin or isospin-symmetry breaking. We shall, however,
assume that the matrix elements of both $1/m_b^3$ and $1/m_b^4$ operators respect
isospin.
The ratio of lifetimes of $B_s$ and $B_d$ mesons involves a breaking of $U$-spin symmetry,
so the matrix elements of dimension-6 operators could differ by about $30\%$. 
We shall introduce different $B-$ and $\epsilon$-parameters to describe $B_s$ and
$B_d$ lifetimes. 

\begin{figure}[t]
\centerline{\epsfig{file=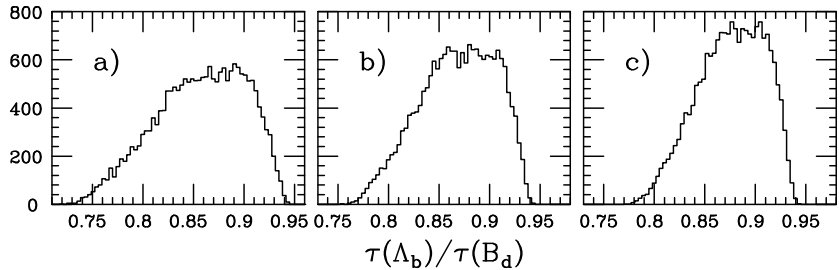, angle=90, width=11cm}}
\centerline{\parbox{12cm}{\caption{\label{fig:randLam}
Histograms showing the random distributions around the central values
of the $f_{B_q}$, $m_b$, $B$, $\delta$, $\epsilon$$, \beta_i$ 
and $\widetilde{\beta}_i$ parameters 
contributing to $\tau(\Lambda_b)/\tau(B_d)$.
Three histograms are shown for the scales $\mu = m_b/2$ (a), $\mu =
m_b$ (b), and $\mu = 2\, m_b$ (c).}}}
\end{figure}

\begin{figure}[t]
\centerline{\epsfig{file=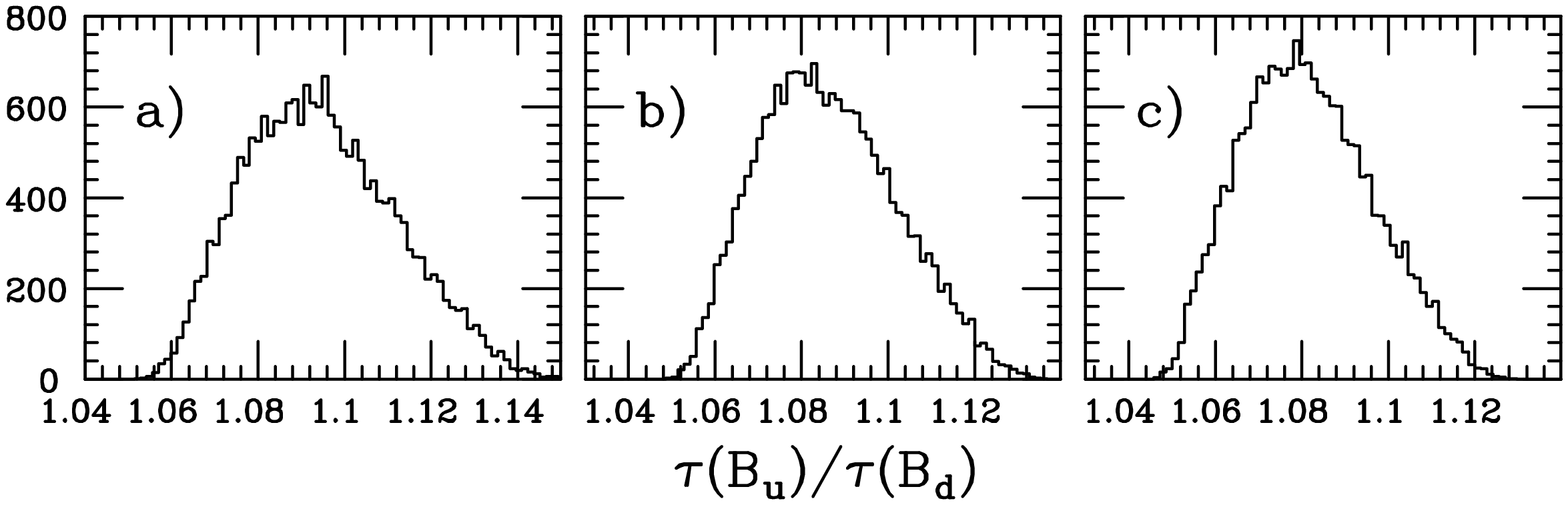, angle=90, width=11cm}}
\centerline{\parbox{12cm}{\caption{\label{fig:randU}
Same as Fig.~\ref{fig:randLam} for $\tau(B_u)/\tau(B_d)$.}}}
\end{figure}

\begin{figure}[t]
\centerline{\epsfig{file=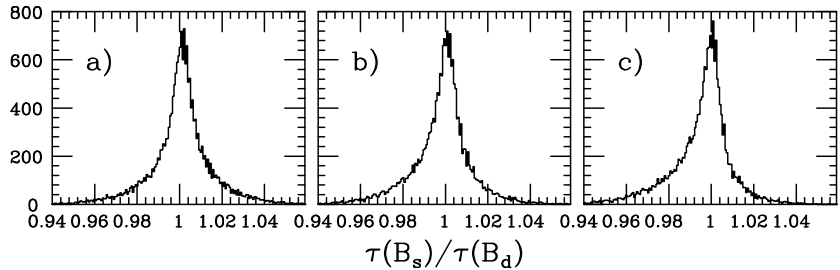, angle=90, width=11cm}}
\centerline{\parbox{12cm}{\caption{\label{fig:randS}
Same as Fig.~\ref{fig:randLam} for $\tau(B_s)/\tau(B_d)$.}}}
\end{figure}

In order to obtain numerical estimate of the effect of $1/m_b$ corrections to
spectator effects, we adopt the statistical approach for
presenting our results and generate 20000-point probability distributions of the 
ratios of lifetimes obtained by randomly varying the parameters 
describing matrix elements within a $\pm 30\%$ interval around 
their ``factorization'' values\cite{Gabbiani:2003pq}, for three different 
scales $\mu$. The decay constants $f_{B_q}$ and b-quark pole mass $m_b$ are 
taken to vary within $1\sigma$ interval indicated above.
The results are presented in Figs.~\ref{fig:randLam},~\ref{fig:randU},
and \ref{fig:randS}. These figures represent our main result.
We also performed studies of convergence of $1/m_b$ expansion by computing a set of
$1/m_b^2$-corrections to spectators effects and estimating their size
in factorization\cite{Gabbiani:2003pq}. The expansion appears to be well-convergent 
for the $b$-flavored hadrons. Due to the relative smallness of $m_c$ (and thus
vicinity of the sector of QCD populated by the light quark resonances\cite{charm}) it 
is not clear that the application of these findings to charmed hadrons will produce 
quantitative, rather than qualitative results.

\section{Conclusions}\label{Conclusions}

We computed subleading $1/m_b$ and $1/m_b^2$ corrections to the spectator effects
driving the difference in the lifetimes of heavy mesons and baryons. Thanks to the same
$16\pi^2$ phase-space enhancement as $1/m_b^3$-suppressed spectator effects, these
corrections constitute the most important set of $1/m_b^4$-suppressed corrections.

The main result of this talk are Figs.~\ref{fig:randLam},~\ref{fig:randU}, and
\ref{fig:randS}, which represent the effects of subleading spectator effects on
the ratios of lifetimes of heavy mesons and baryons.
We see that subleading corrections to spectator effects affect the ratio of heavy meson
lifetimes only modestly, at the level of a fraction of a percent. On the other hand,
the effect on the $\Lambda_b$-$B_d$ lifetime ratio is quite substantial, at the level of
$-3\%$. This can be explained by the partial cancellation of WS and PI effects in
$\Lambda_b$ baryon and constructive interference of $1/m_b$ corrections to the spectator
effects. 

There is no theoretically-consistent way to translate the histograms of 
Figs.~\ref{fig:randLam},~\ref{fig:randU}, and \ref{fig:randS} into numerical predictions
for the lifetime ratios. As a useful estimate it is possible to fit the histograms to gaussian 
distributions and extract theoretical predictions for the mean values and deviations of the 
ratios of lifetimes. Predictions obtained this way should be treated with care, as it is not 
expected that the theoretical predictions are distributed according to the gaussian 
distribution. This being said, we proceed by fitting the distributions to gaussians 
and, correcting for the small scale uncertainty, 
extract the ratios $\tau(B_u)/\tau(B_d)= 1.09 \pm 0.03$, 
$\tau(B_s)/\tau(B_d)= 1.00 \pm 0.01$, and $\tau(\Lambda_b)/\tau(B_d)= 0.87
\pm 0.05$. This brings the experimental and theoretical ratios of baryon and meson lifetimes
into agreement.

I would like to thank F.~Gabbiani and A.~I.~Onishchenko for collaboration on
this project, and N.~Uraltsev and M.~Voloshin for helpful discussions.
It is my pleasure to thank the organizers for the invitation to this wonderfully
organized workshop.


\end{document}